\documentclass{amsart}

\usepackage{amsmath,amsfonts}
\usepackage{amsthm}
\usepackage{tikz}
\usetikzlibrary{decorations.markings}
\usetikzlibrary{arrows}
\usepackage{graphicx}
\usepackage{float}

\newtheorem{thm}{Theorem}

\begin{document}

\title{An algebraic approach to the minimum-cost multi-impulse orbit transfer problem}

\author{M.~Avenda\~no}
\address{Centro Universitario de la Defensa, Ctra. Huesca s/n, Zaragoza 50090, Spain}
\email{avendano@unizar.es}

\author{V.~Mart\'{\i}n-Molina}
\address{Departamento de Did\'{a}ctica de las Matem\'{a}ticas, Facultad de Ciencias de la Educaci\'on, c/ Pirotecnia s/n, Universidad de Sevilla, Sevilla 41013, Spain}
\email{veronicamartin@us.es}

\author{J.~Mart\'\i n-Morales}
\address{Centro Universitario de la Defensa, Ctra. Huesca s/n, Zaragoza 50090, Spain}
\email{jorge@unizar.es}

\author{J.~Ortigas-Galindo}
\address{Instituto de Educaci\'on Secundaria \'Elaios, c/ Andador Pilar Cuartero 3, Zaragoza 50018, Spain}
\email{jortigas@educa.aragon.es}

\begin{abstract}
We present a purely algebraic formulation (i.e. polynomial equations only) of the minimum-cost multi-impulse orbit transfer problem without time constraints, while keeping all the
variables with a precise physical meaning. We apply general algebraic techniques to solve
these equations (resultants, Gr\"obner bases, etc.) in several situations of practical
interest of different degrees of generality. For instance, we provide a proof of the
optimality of the Hohmann transfer for the minimum fuel 2-impulse circular to circular
orbit transfer problem, and we provide a general formula for the optimal 2-impulse in-plane
transfer between two rotated elliptical orbits under a mild symmetry assumption on the
two points where the impulses are applied (which we conjecture that can be removed).
\end{abstract}

\maketitle

\section{Introduction}\label{sec-intro}

Since the start of the space age with the first launch of a satellite
to space (Sputnik in $1957$), the interest in the study of space maneuvers that use the available resources like time and fuel efficiently has been growing steadily. In many real life cases, a satellite can serve multiple purposes, requiring for that a change of its orbit.
Some maneuvering is also needed during the initial launch of a satellite, or when a spare satellite
has to  be brought to its intended orbit.

Maneuvering a satellite can be done in two different ways: continuous
thrust or a sequence of instantaneous and discrete impulses. This paper
focuses on the latter.

The orbit transfer problem with a fixed time of flight was studied by Lambert, who provided a solution in the case of two impulses. For a discussion of this problem, see \cite{Val97}.

However, the scarcest resource in space is fuel, since it represents a load on the spacecraft that cannot be too large to avoid launching problems and to reduce costs. For this reason, we focus our attention on
the minimum fuel transfer problem with unconstrained time. Using the well-known
Tsiolkovsky rocket equation, we consider the sum of the individual impulses (difference between velocities before and after the thrust is applied) as the cost function. This usually appears in the literature as $\Delta v$.

In the case of a transfer between two circular coplanar orbits, Hohmann gave an explicit solution with two impulses in~\cite{Hoh25}, which was later proven optimal analytically by Barrar~\cite{Bar63}. For the case with three impulses, Hoelker and Silber~\cite{HS59} have shown that a bi-elliptical transfer has a lower fuel requirement than the Hohmann transfer for some special initial and final orbits. Roth~\cite{Rot67} extended the notion of bi-elliptical transfer to the case of two inclined orbits.

In this paper, we provide a detailed study of transfers
between two circular orbits, including out-of-plane maneuvers and also
the possibility that the initial and final angular momentum point in
opposite directions. In all cases, we have proven algebraically that the Hohmann transfer is optimal for two impulses.

Another problem of interest is transferring a satellite between predetermined points in the initial and final orbits. This situation was
studied by Avenda\~no and Mortari in~\cite{AM10}, where they provided a closed-form solution. Here, we reobtain this solution
algebraically, applying a much more efficient method. Previous attempts to solve this problem involved the use of iterative methods or an equation that needs to be solved numerically (see \cite{AP63,GD69,PB96,Sch97,SP06}).

The last problem we study is the optimal transfer between two identical
ellipses that are coplanar and rotated a fixed angle. This case was
studied numerically by Bender in~\cite{Ben62}. However, we provide an
algebraic solution, which is fully explicit under a mild symmetry assumption.


This paper is organized as follows. In Section~\ref{sec-kepler} we have compiled all the equations of Celestial Mechanics that we will need in the paper. In Section~\ref{sec-modelo}, we present an algebraic approach to the multi-impulse minimum-cost orbit transfer problem. We have put special emphasis in explaining the physical meaning of all the variables involved. Three kinds of problems are studied: point to point, point to orbit and orbit to orbit. We also consider two possible cost functions.

In Section~\ref{sec-am}, we present our solution to the general point-to-point problem with two impulses and cost function as in \cite{AM10}.
In Section~\ref{sec-hohmann}, we provide a solution to a generalized version of the Hohmann transfer where out-of-plane maneuvers are allowed.

Finally, the orbit-to-orbit problem between identical and coplanar orbits is studied in Section~\ref{sec-two-rot}. The solution we obtain requires solving a large system of polynomial equations, which can be solved explicitly if we assume a symmetry condition. We have done extensive numerical tests showing that the symmetry condition is always satisfied in them.

\section{Keplerian motion}\label{sec-kepler}

The motion of a particle in a Keplerian gravitational force field is given by the
solution of the second-order differential equation
\begin{equation}\label{eq1}
  \vec{r}(t_0)=\vec{r}_0,\;
  \dot{\vec{r}}(t_0)=\vec{v}_0,\;
  \ddot{\vec{r}}=-\frac{\mu}{|\vec{r}|^3}\vec{r},
\end{equation}
where $\mu>0$ is the standard gravitational parameter of the field, $\vec{r}_0$ and
$\vec{v}_0$ are the initial position and velocity, and $\vec{r}(t)$ is the position
of the particle as a function of time. For any solution of Eq.\eqref{eq1}, the angular
momentum vector
\begin{equation*}
  \vec{h}=\vec{r}\times\dot{\vec{r}},
\end{equation*}
the eccentricity vector
\begin{equation}\label{eq3}
  \vec{e}=\frac{\dot{\vec{r}}\times\vec{h}}{\mu}-\frac{\vec{r}}{|\vec{r}|},
\end{equation}
and the total energy
\begin{equation*}
  E=\frac{|\dot{\vec{r}}|^2}{2}-\frac{\mu}{|\vec{r}|}
\end{equation*}
are constants with respect of time~\cite[Ch.~8.3]{SJ03}. The vectors $\vec{h}$ and $\vec{e}$ are always
orthogonal, i.e.
\begin{equation*}
  \vec{h}\cdot\vec{e}=0,
\end{equation*}
and any pair of mutually orthogonal vectors $\vec{h}$ and $\vec{e}$ can be obtained
for some initial conditions $\vec{r_0}$ and $\vec{v_0}$. Moreover, the total energy
satisfies
\begin{equation*}
  1-|\vec{e}|^2 = -\frac{2E|\vec{h}|^2}{\mu^2},
\end{equation*}
so its value can be determined from $\vec{h}$ and $\vec{e}$ alone when $\vec{h} \neq \vec{0}$.

The angular momentum is always orthogonal to $\vec{r}$, i.e.
\begin{equation}\label{eq4x}
  \vec{r}\cdot\vec{h} = 0,
\end{equation}
so the motion is planar.
Besides, it follows from Eq.\eqref{eq3} that
\begin{equation}\label{eq4}
  |\vec{r}|+\vec{e}\cdot\vec{r}
  =\frac{\dot{\vec{r}}\times\vec{h}}{\mu}\cdot\vec{r}
  =\frac{|\vec{h}|^2}{\mu}
\end{equation}
is also constant, which is the implicit equation of a conic with one focus at the
origin, eccentricity $e=|\vec{e}|$ and semilatus rectum $p=|\vec{h}|^2/\mu$, when
$p\neq 0$. In the case $e>1$, i.e. when the conic is a hyperbola, the equation
describes only the branch in which the particle is moving. The degenerate case $p=0$ will be discussed later in this section.
Finally, multiplying Eq.\eqref{eq3} by $\vec{h}$ and moving some terms, we obtain an expression for the velocity of the particle at any given position:
\begin{equation}\label{eq5}
  \dot{\vec{r}}=\mu\frac{\vec{h}}{|\vec{h}|^2}\times\left(\vec{e}+\frac{\vec{r}}{|\vec{r}|}\right) \, .
\end{equation}

It is important to note that, when the orbit is an ellipse ($e<1$), any point
$\vec{r}$ that satisfies Eq.~\eqref{eq4x} and Eq.~\eqref{eq4} will be visited by the
particle at some time $t\geq t_0$ since the motion is periodic. However, this is
not true when $e\geq 1$ because in the case of a parabolic ($e=1$) or hyperbolic ($e>1$) trajectory only the points satisfying the extra condition
\begin{equation*}
  (\vec{r}\times\vec{r_0})\cdot\vec{h}\leq 0
\end{equation*}
will be visited.

As we mentioned above, the case $p=0$ needs to be discussed separately. Here we
have $\vec{h}=\vec{0}$, so $\vec{r}_0$ is parallel to $\vec{v}_0$. The eccentricity
vector $\vec{e}=-\frac{\vec{r}}{|\vec{r}|}$ is constant, so the trajectory is
contained in the line through the origin with direction $\vec{e}$.
There are two possible cases: either the initial velocity is high enough to escape the gravitational attraction of the field or the particle will first move in the direction $\vec{v}_0$ until a point where its velocity becomes zero and then come back towards the origin, thus entering in a periodic motion.

To avoid the extra complexity needed to handle parabolic and hyperbolic motions,
as well as the degenerate case $p=0$ described above, we restrict our analysis
to elliptic orbits, $e<1$, and non-degenerate trajectories, $\vec{h}\neq \vec{0}$.

In order to work with polynomial equations, we need to remove the divisions, the square roots and the constant $\mu$ from some of the equations above, so we introduce the vectors
\begin{equation*}
  \hat{\vec{r}}=\frac{\vec{r}}{|\vec{r}|},\;
  \vec{w}=\frac{\dot{\vec{r}}}{\sqrt{\mu}},\;
  \vec{l}=\sqrt{\mu}\frac{\vec{h}}{|\vec{h}|^2},\;
  \vec{s}=\vec{l}\times\vec{e} \, .
\end{equation*}
Note that $\vec{l}$ and $\vec{s}$ are orthogonal, i.e.
\begin{equation}\label{eq9x}
  \vec{l}\cdot\vec{s}=0,
\end{equation}
and that the angular momentum and eccentricity vectors can be simply recovered as
\begin{equation*}
  \vec{h}=\sqrt{\mu}\frac{\vec{l}}{|\vec{l}|^2},\;
  \vec{e}=\frac{\vec{s}\times\vec{l}}{|\vec{l}|^2}.
\end{equation*}
Of course, the case $\vec{l}=\vec{0}$ has to be excluded, and the condition $e<1$
translates into $|\vec{s}|<|\vec{l}|$.

Any unit vector $\hat{\vec{r}}$ orthogonal to $\vec{l}$, i.e.
\begin{equation}\label{eq10}
  \hat{\vec{r}}\cdot\hat{\vec{r}}=1,\;
  \hat{\vec{r}}\cdot\vec{l}=0,
\end{equation}
determines a point on the orbit. The exact location can be obtained from Eq.~\eqref{eq4},
\begin{equation}\label{eq10x}
  \frac{1}{|\vec{r}|}=|\vec{l}|^2 + (\vec{s}\times\vec{l})\cdot\hat{\vec{r}},
\end{equation}
and the velocity of the particle at that point is, according to Eq.~\eqref{eq5},
\begin{equation}\label{eq11}
  \vec{w}=\vec{s}+\vec{l}\times\hat{\vec{r}} \, .
\end{equation}
Finally, note that in the case of elliptic orbits ($|\vec{s}|<|\vec{l}|$), the right-hand side of Eq.~\eqref{eq10x} is always positive, so no extra inequalities are needed
to guarantee a valid value of $|\vec{r}|^{-1}$.


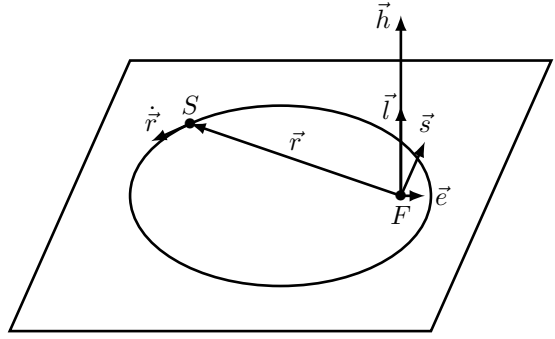
\begin{figure}[H]
\centering
\begin{tikzpicture}[line width=1pt,scale=0.4]
  \coordinate (F) at (0,0);    
  \coordinate (O) at (-4,0);   
  \coordinate (S) at (-7,2.4); 

  \node [] at (F) {\textbullet};
  \node [below] at (F) {$F$};
  \node [] at (S) {\textbullet};
  \node [above] at (S) {$S$};

  \draw (O) circle [x radius=5cm, y radius=3cm,];

  \path[draw] (5,4.5) -- (1,-4.5) -- (-13,-4.5) -- (-9,4.5)-- cycle;

  \draw[-latex,] (F)--(S) node[midway,above] {$\vec{r}$};
  \draw[-latex,] (F)-- +(0,3) node[ left] {$\vec{l}$};
  \draw[-latex,] (F)-- +(0,6) node[left] {$\vec{h}$};
  \draw[-latex,] (F)-- +(0.8,0) node[right] {$\vec{e}$};
  \draw[-latex,] (F)-- +(0.8,1.8) node[above] {$\vec{s}$};
  \draw[-latex,] (S)-- +(-1.3,-0.585) node[above] {$\dot{\vec{r}}$};
\end{tikzpicture}
\caption{Ellipse with focus $F$ and a satellite $S$ on it.}
\end{figure}

\section{Multi-impulse orbit transfers}\label{sec-modelo}

An $n$-impulse orbit transfer is represented algebraically by the vectors
\[
  \vec{l}_0,\vec{l}_1,\ldots,\vec{l}_n,
  \vec{s}_0,\vec{s}_1,\ldots,\vec{s}_n,
  \hat{\vec{r}}_0,\hat{\vec{r}}_1,\ldots,\hat{\vec{r}}_{n-1}
\]
where the pair $(\vec{l}_i,\vec{s}_i)$ determines the $i$-th orbit and $\hat{\vec{r}}_i$ corresponds to the point where the impulse is applied to change from the $i$-th to the $(i+1)$-th orbit. These vectors, according
to Eqs.\eqref{eq9x},\eqref{eq10} and \eqref{eq10x}, are constrained by
\begin{align}
  &\vec{l}_i\cdot\vec{s}_i=0,\\
  &\vec{l}_i\neq 0, \\
  &|\vec{s}_i|<|\vec{l}_i|,
\end{align}
for $i=0,\ldots,n$, and
\begin{align}
  \vec{l}_i\cdot\hat{\vec{r}}_i&=0,\\
  \vec{l}_{i+1}\cdot\hat{\vec{r}}_i&=0,\\
  |\hat{\vec{r}}_i|^2&=1,\\
  |\vec{l}_i|^2+(\vec{s}_i\times\vec{l}_i)\cdot\hat{\vec{r}}_i,
    &=|\vec{l}_{i+1}|^2+(\vec{s}_{i+1}\times\vec{l}_{i+1})\cdot\hat{\vec{r}}_i,
\end{align}
for all $i=0,\ldots,n-1$. Conversely, any sequence of vectors satisfying all these
restrictions represents a valid $n$-impulse transfer.
Moreover, all the equations are invariant under rotation by a fixed angle and rescaling of the $\vec{l}_i$ and $\vec{s}_i$.

The following table shows the number of unknowns and algebraic equations that
define the configuration space for each type of $n$-impulse transfer ($n\geq 2$).
\begin{center}
\begin{tabular}{|c||c|c|c|c|c|c|}
\hline
                & \multicolumn{2}{c|}{3d-transfer} &  \multicolumn{2}{c|}{2d-transfer} \\
                & \#unknowns  & \#equations & \#unknowns  & \#equations  \\
\hline
Point to Point  & $9n-12$ & $5n-5$ & $5n-7$  & $2n-2$  \\
Point to Orbit  & $9n-9$  & $5n-3$ & $5n-5$  & $2n-1$  \\
Orbit to Orbit  & $9n-6$  & $5n-1$ & $5n-3$  & $2n$    \\
\hline
\end{tabular}
\end{center}
In the orbit to orbit problem, the vectors $\vec{l}_0$, $\vec{s}_0$, $\vec{l}_n$, $\vec{s}_n$
are given and the remaining variables are considered unknowns. In the point
to orbit, the initial point is given, so $\hat{\vec{r}}_{0}$ is also known, thus
reducing the number of unknowns (and equations). Finally, in the point to point
problem, the final point is also given, i.e.~$\hat{\vec{r}}_{n-1}$ is known.

The two-dimensional version of these problems considers all orbits in the $z=0$ plane, so
$\vec{l}_i=(0,0,l_{iz})$, $\vec{s}_i=(s_{ix}, s_{iy}, 0)$ and
$\hat{\vec{r}}_i=(x_i,y_i,0)$, hence the reduced number of variables
and equations needed to handle them.

The velocities at the points $\hat{\vec{r}}_i$, immediately before and after the impulse is applied, are written $\vec{w}_i$ and $\vec{w}_i^{\ast}$, respectively. It follows from Eq.\eqref{eq11} that
\begin{align}
  \vec{w}_i         &=\vec{s}_i+\vec{l}_i\times\hat{\vec{r}}_i, \label{eq-wi}\\
  \vec{w}_i^{\ast}  &=\vec{s}_{i+1}+\vec{l}_{i+1}\times\hat{\vec{r}}_i. \label{eq-wi*}
\end{align}
The cost (fuel-wise) of such a transfer is proportional to the sum of
$\Delta_i=|\vec{w}_i-\vec{w}_i^{\ast}|$, denoted hereafter by $f_1$.
To avoid the square roots that are implicitly present in $\Delta_i$, we also consider a cost function $f_2$ which is the sum of the squares of the $\Delta_i$:
\begin{equation*}
  f_1=\sum_{i=0}^{n-1}|\vec{w}_i-\vec{w}_i^{\ast}|,\quad
  f_2=\sum_{i=0}^{n-1}|\vec{w}_i-\vec{w}_i^{\ast}|^2 \, .
\end{equation*}

If the vectors $\vec{l}_i$ and $\vec{s}_i$ are rescaled by a factor $c$, then $f_1$ and $f_2$ are multiplied by a factor $|c|$ and $|c|^2$, respectively.

When the cost function $f_1$ is used, the trick to avoid the square roots consists of considering $\Delta_i$ as a variable, redefining the cost function as
\begin{equation*}
  f_1=\sum_{i=0}^{n-1}\Delta_i
\end{equation*}
and adding the algebraic equations
\begin{equation*}
  \Delta_i^2 =
  |\vec{s}_i-\vec{s}_{i+1}|^2+|\vec{l}_i-\vec{l}_{i+1}|^2+
     2((\vec{s}_i-\vec{s}_{i+1})\times(\vec{l}_i-\vec{l}_{i+1}))\cdot\hat{\vec{r}}_i \, ,
\end{equation*}
for $i=0,\ldots,n-1$. The last equations can be obtained by substituting Eqs.\eqref{eq-wi} and \eqref{eq-wi*} into the definition of $\Delta_i$ \, :
\begin{align*}
\Delta_i^2 &= |\vec{w}_i-\vec{w}_i^{\ast}|^2 = |\vec{s}_i+\vec{l}_i \times \hat{\vec{r}}_i
  -\vec{s}_{i+1}-\vec{l}_{i+1} \times \hat{\vec{r}}_i  |^2 \\
  &=|\vec{s}_i-\vec{s}_{i+1}|^2+|\vec{l}_i-\vec{l}_{i+1}|^2+
     2((\vec{s}_i-\vec{s}_{i+1})\times(\vec{l}_i-\vec{l}_{i+1}))\cdot\hat{\vec{r}}_i \, .
\end{align*}

At this point we have a classical problem of constrained minimization, which we approach with
Lagrange multipliers.

\begin{thm}[Lagrange multipliers]\label{thm-lm}
  Let $q,q_1,\ldots,q_m:\mathbb{R}^k\to\mathbb{R}$ in $\mathcal{C}^\infty$ and $p\in\mathbb{R}^k$
  a common zero of $q_1,\ldots,q_m$ be such that the vectors $\nabla q_1(p),\ldots,\nabla q_m(p)$
  are linearly independent. Then $p$ is a local extremum of $q$ on the manifold defined by
  $\{ q_1=\cdots=q_m=0 \}$ if and only if there exists $\lambda_1,\ldots,\lambda_m\in\mathbb{R}$
  such that $\nabla q(p)=\lambda_1\nabla q_1(p)+\cdots+\lambda_m\nabla q_m(p)$.
\end{thm}

In our case, we have an algebraic variety $V=\{ q_1=\ldots=q_m=0 \}\subseteq\mathbb{R}^k$, defined
by polynomials $q_1,\ldots,q_m\in\mathbb{R}[x_1,\ldots,x_k]$, and another polynomial function $q$
that we want to minimize on $V$. In order to apply Theorem~\ref{thm-lm},
we need to exclude first the points where $\nabla q_1(p),\ldots,\nabla q_m(p)$ are not
linearly independent, which is, by definition, the set of singular points $V^\ast$ of $V$. Computationally, $V^\ast$ is the set of points of $V$ where all $m\times m$ minors of the matrix
$[\partial q_i/\partial x_j]_{1\leq i\leq m,1\leq j\leq k}$ have zero determinant:
\[
V^\ast = \left\{p\in\mathbb{R}^k\,:\,q_1=\cdots=q_m=0\;\wedge\;
            \left| \frac{\partial (q_1,\ldots,q_m)}{\partial (x_{j_1},\ldots,x_{j_m})}\right|=0,
            \forall J=\{j_1, \ldots, j_m \}\subseteq\{1,\ldots,k\}
         \right\} \, .
\]
These points have to be considered critical points (i.e. they are potential local extrema) and have to be evaluated separately.

On the remaining points, $V\setminus V^\ast$, the local extrema can be
found directly by Theorem~\ref{thm-lm}, solving the system of $m+k$ equations $q_1=\cdots=q_m=0$
and $\nabla q=\lambda_1\nabla q_1+\cdots+\lambda_m\nabla q_m$ in the $m+k$ unknowns $x_1,\ldots,x_k,\lambda_1,\ldots,\lambda_m\in\mathbb{R}$, and disregarding the solutions with $(x_1,\ldots,x_k)\in V^\ast$.
Removing these solutions is actually not needed since they are always critical points. The set of
solutions of the $m+k$ equations described above and the set of all critical points of $q$ are denoted
$V_q$ and $V_q^{crit}$, respectively:
\begin{equation}\label{eq-alfa}
\begin{aligned}
  V_q &= \left\{ (x_1,\ldots,x_k,\lambda_1,\ldots,\lambda_m)\in\mathbb{R}^{m+k}\,:\,
          {{q_1=\cdots=q_m=0} \;\wedge\; {\nabla q = \lambda_1\nabla q_1+\cdots+\lambda_m\nabla q_m}}
        \right\},  \\
  V_q^{crit}&=V^\ast\cup\pi_k\left(V_q\right)\subseteq\mathbb{R}^k,
\end{aligned}
\end{equation}
where $\pi_k : \mathbb{R}^{m+k} \to \mathbb{R}^{k}$ is the projection onto the first $k$ coordinates.

Including the Lagrange multipliers, and the extra variables $\Delta_i$ for $i=0,\ldots,n-1$
when minimizing $f_1$ instead of $f_2$, we get the following total number of unknowns (which is
equal to the number of equations):
\begin{center}
\begin{tabular}{|c||c|c|c|c|}
\hline
                & \multicolumn{2}{c|}{$\min(f_1)$} & \multicolumn{2}{c|}{$\min(f_2)$} \\
                & 3d-transfer  & 2d-transfer    & 3d-transfer & 2d-transfer \\
\hline
Point to Point  & $16n-17$     & $9n-9$         & $14n-17$    & $7n-9$  \\
Point to Orbit  & $16n-12$     & $9n-6$         & $14n-12$    & $7n-6$  \\
Orbit to Orbit  & $16n-7$      & $9n-3$         & $14n-7$     & $7n-3$  \\
\hline
\end{tabular}
\end{center}

By using standard linear algebra, it is possible to eliminate all the Lagrange multipliers:
\begin{equation} \label{eq-beta}
  V_q^{crit}=V^\ast\cup
     \left\{
        \begin{aligned}
        &q_1=\cdots=q_m=0 \\
        &\left|
           \frac{\partial (q,q_1,\ldots,q_m)}{\partial (x_{j_1},\ldots,x_{j_{m+1}})}\right|=0,
           \;\;\forall J=\{j_1,\ldots,j_{m+1}\} \subseteq\{1,\ldots,k\}
        \end{aligned}
     \right\} \, .
\end{equation}
The expression above shows that $V_q^{crit}$ can be written as the union of two algebraic
varieties in $\mathbb{R}^k$.

\section{Minimum $\Delta v^2$ Lambert problem}\label{sec-am}

In this problem, the vectors $\vec{r}_0$, $\vec{r}_1$, $\vec{w}_0$,
$\vec{w}^{\ast}_1$ are known, from which the vectors $\vec{l}_0$, $\vec{l}_2$,
$\vec{s}_0$, $\vec{s}_2$ can be computed directly. The unknowns are $\vec{l}_1$ and $\vec{s}_1$, from which we can deduce $\vec{w}_0^*$ and $\vec{w}_1$. There are two cases, depending on whether $\vec{r}_0$ and $\vec{r}_1$ are linearly independent or not.

In the first case, we can assume without loss of generality that $\vec{r}_0$
and $\vec{r}_1$ both lie on the $xy$-plane, so the unknowns can be written
$\vec{l}_1=(0,0,l_{1z})$ and $\vec{s}_1=(s_{1x},s_{1y},0)$. We can further assume
that $\hat{\vec{r}}_0=(1,0,0)$ and $\hat{\vec{r}}_1=(x_1,y_1,0)$ with
$y_1\neq 0$ and $x_1^2+y_1^2=1$. Finally, if we define $k_0=|\vec{r}_0|^{-1}$
and $k_1=|\vec{r}_1|^{-1}$, we obtain the following two restrictions:
\begin{align}
  q_1 &:=l_{1z}^2 + l_{1z}s_{1y} - k_0 =0, \label{am1} \\
  q_2 &:=l_{1z}^2 + l_{1z}(x_1s_{1y}-y_1s_{1x}) -k_1=0.
\end{align}

From Eq.~\eqref{eq11}, the velocities $\vec{w}^{\ast}_0$ and $\vec{w}_1$ are
\begin{align}
  \vec{w}^{\ast}_0 &= \vec{s}_1+\vec{l}_1\times\hat{\vec{r}}_0=(s_{1x},s_{1y}+l_{1z},0), \\
  \vec{w}_1 &= \vec{s}_1+\vec{l}_1\times\hat{\vec{r}}_1=(s_{1x}-l_{1z}y_1, s_{1y}+l_{1z}x_1,0),\label{am4}
\end{align}
and the impulses $\Delta_0$ and $\Delta_1$ are given by
\begin{align}
  \Delta_0 &=| \vec{w}_0^* - \vec{w}_0 | =  |(s_{1x}-w_{0x},s_{1y}+l_{1z}-w_{0y},-w_{0z})|, \\
  \Delta_1 &= | \vec{w}_1^* - \vec{w}_1 | =
  |(w_{1x}^{\ast}-s_{1x}+l_{1z}y_1, w_{1y}^{\ast}-s_{1y}-l_{1z}x_1,w_{1z}^{\ast})|,
\end{align}
so the cost function $q=f_2=\Delta_0^2+\Delta_1^2$ is
\[
q:= (s_{1x}-w_{0x} )^2 +(s_{1y}+l_{1z}-w_{0y})^2 +(w_{0z})^2
+(w_{1x}^{\ast}-s_{1x}+l_{1z}y_1)^2 +(w_{1y}^{\ast}-s_{1y}-l_{1z}x_1)^2 +(w_{1z}^*)^2.
\]
We compute the critical points of $q$ using Eq.~\eqref{eq-beta}. In this case, $V^*=\emptyset$ because
\[
\left|
  \frac{\partial (q_1,q_2)}{\partial (s_{1x},s_{1y})}
\right|=l_{1z}^2 y_1=0
\]
is impossible since $l_{1z}\neq0$ and $y_1\neq0$. Therefore, Eq.~\eqref{eq-beta} reduces to
\begin{equation}\label{eq-27}
  V_q^{crit}=
     \left\{
        q_1=q_2=
        \left|
             \frac{\partial (q,q_1,q_2)}{\partial (s_{1x},s_{1y},l_{1z})}
        \right|=0
     \right\} \, .
\end{equation}

We will solve Eqs.~\eqref{eq-27} using the technique explained in \cite[Ch.2]{CLO05}. In order to do that, we computed the Gr\"obner basis of $V_q^{crit}$ in
the polynomial ring $K[s_{1x},s_{1y},l_{1z}]$ over the field
of fractions
$K={\rm Frac}\left(\mathbb{Q}[k_0,k_1,x_1,y_1,\vec{w}_0,\vec{w}^{\ast}_1]/\langle x_1^2+y_1^2-1\rangle\right)$
with respect to the lexicographic monomial order $s_{1x}>s_{1y}>l_{1z}$,
obtaining $I=\langle p_1,p_2,p_3\rangle$, where
\begin{align*}
  p_1 &= k_0 y_1\cdot s_{1x}-(k_0 x_1-k_1)\cdot s_{1y}-(k_0-k_1)\cdot l_{1z} \\
  p_2 &= (2(k_0^2-k_1^2)^2+8k_0^{2}k_1^{2}y_1^{2})\cdot s_{1y} \\
      &\quad +(4k_0^{3}x_1+2k_0^{3}y_1^{2}-4k_0^{3}+4k_0^{2}k_1 x_1 y_1^{2}-8k_0^{2}k_1 x_1-8k_0^{2}k_1 y_1^{2}+8k_0^{2}k_1 \\
      &\hspace*{1cm}+4k_0 k_1^{2}x_1+2k_0 k_1^{2}y_1^{2}-4k_0 k_1^{2})\cdot l_{1z}^{3} \\
      &\quad +(k_0^{3}x_1 y_1^{2}w_{1y}^{\ast}-k_0^{3}x_1 y_1 w_{0x}-k_0^{3}x_1 y_1 w_{1x}^{\ast}-k_0^{3}y_1^{3}w_{1x}^{\ast}-k_0^{3}y_1^{2}w_{1y}^{\ast}+k_0^{3}y_1 w_{0x}\\
      &\hspace*{1cm}+k_0^{3}y_1 w_{1x}^{\ast}-2k_0^{2}k_1 x_1 y_1^{3}w_{1x}^{\ast}-2k_0^{2}k_1 x_1 y_1^{2}w_{1y}^{\ast}+2k_0^{2}k_1 x_1 y_1 w_{0x}+2k_0^{2}k_1 x_1 y_1 w_{1x}^{\ast}\\
      &\hspace*{1cm}-2k_0^{2}k_1 y_1^{4}w_{1y}^{\ast}+2k_0^{2}k_1 y_1^{3}w_{0x}+2k_0^{2}k_1 y_1^{3}w_{1x}^{\ast}+2k_0^{2}k_1 y_1^{2}w_{1y}^{\ast}-2k_0^{2}k_1 y_1 w_{0x}\\
      &\hspace*{1cm}-2k_0^{2}k_1 y_1 w_{1x}^{\ast}+k_0 k_1^{2}x_1 y_1^{2}w_{1y}^{\ast}-k_0 k_1^{2}x_1 y_1 w_{0x}-k_0 k_1^{2}x_1 y_1 w_{1x}^{\ast}-k_0 k_1^{2}y_1^{3}w_{1x}^{\ast}\\
      &\hspace*{1cm}-k_0 k_1^{2}y_1^{2}w_{1y}^{\ast}+k_0 k_1^{2}y_1 w_{0x}+k_0 k_1^{2}y_1 w_{1x}^{\ast})\cdot l_{1z}^{2} \\
      &\quad +(2k_0^{4}+8k_0^{2}k_1^{2}y_1^{2}-4k_0^{2}k_1^{2}+2k_1^{4})\cdot l_{1z} \\
      &\quad -(k_0^{4}x_1 y_1 w_{0x}+k_0^{4}x_1 y_1 w_{1x}^{\ast}+k_0^{4}y_1^{2}w_{0y}+k_0^{4}y_1^{2}w_{1y}^{\ast}+2k_0^{3}k_1 x_1 y_1^{2}w_{0y}+2k_0^{3}k_1 x_1 y_1^{2}w_{1y}^{\ast}\\
      &\hspace*{1cm}-2k_0^{3}k_1 y_1^{3}w_{0x}-2k_0^{3}k_1 y_1^{3}w_{1x}^{\ast}+k_0^{3}k_1 y_1 w_{0x}+k_0^{3}k_1 y_1 w_{1x}^{\ast}-k_0^{2}k_1^{2}x_1 y_1 w_{0x}\\
      &\hspace*{1cm}-k_0^{2}k_1^{2}x_1 y_1 w_{1x}^{\ast}+k_0^{2}k_1^{2}y_1^{2}w_{0y}+k_0^{2}k_1^{2}y_1^{2}w_{1y}^{\ast}-k_0 k_1^{3}y_1 w_{0x}-k_0 k_1^{3}y_1 w_{1x}^{\ast}) \\
  p_3 &= 2y_1^{4}\cdot l_{1z}^{4} +(x_1 y_1^{4}w_{1y}^{\ast}-x_1 y_1^{3}w_{0x}+x_1 y_1^{3}w_{1x}^{\ast}-y_1^{5}w_{1x}^{\ast}+y_1^{4}w_{1y}^{\ast}-y_1^{3}w_{0x}+y_1^{3}w_{1x}^{\ast})\cdot l_{1z}^{3} \\
      &\quad  -(k_0 x_1 y_1^{3}w_{0x}+k_0 x_1 y_1^{3}w_{1x}^{\ast}-2k_0 x_1 y_1^{2}w_{0y}-2k_0 x_1 y_1^{2}w_{1y}^{\ast}-2k_0 x_1 y_1 w_{0x}\\
      &\hspace*{1cm}-2k_0 x_1 y_1 w_{1x}^{\ast}+k_0 y_1^{4}w_{0y}+k_0 y_1^{4}w_{1y}^{\ast}+2k_0 y_1^{3}w_{0x}+2k_0 y_1^{3}w_{1x}^{\ast}-2k_0 y_1^{2}w_{0y}\\
      &\hspace*{1cm}-2k_0 y_1^{2}w_{1y}^{\ast}-2k_0 y_1 w_{0x}-2k_0 y_1 w_{1x}^{\ast}+2k_1 x_1 y_1 w_{0x}+2k_1 x_1 y_1 w_{1x}^{\ast}-k_1 y_1^{3}w_{0x}\\
      &\hspace*{1cm}-k_1 y_1^{3}w_{1x}^{\ast}+2k_1 y_1 w_{0x}+2k_1 y_1 w_{1x}^{\ast})\cdot l_{1z} \\
      &\quad  -(4k_0^{2}x_1-2k_0^{2}y_1^{2}+4k_0^{2}+4k_0 k_1 x_1 y_1^{2}-8k_0 k_1 x_1+8k_0 k_1 y_1^{2}-8k_0 k_1+4k_1^{2}x_1-2k_1^{2}y_1^{2}+4k_1^{2})
\end{align*}
The equation $p_3$ allows one to solve for $l_{1z}$, which can be substituted in $p_2$ to get $s_{1y}$ and, finally, in $p_1$ to obtain $s_{1x}$. This can be done since the leading coefficient in each equation is not zero.

\medskip

Now we deal with the case when the vectors $\vec{r}_0$ and $\vec{r}_1$ are linearly dependent. We can reduce to either $\hat{\vec{r}}_0=\hat{\vec{r}}_1=(1,0,0)$ or  $\hat{\vec{r}}_0=-\hat{\vec{r}}_1=(1,0,0)$. There is no need to use all the machinery
that we developed so far to handle these two degenerate cases. The following discussion
shows how to solve both situations with simple geometric arguments.

In the former case, i.e. $\hat{\vec{r}}_0=\hat{\vec{r}}_1=(1,0,0)$, we must have $k_0=k_1$
and $\vec{w}_0^\ast=\vec{w}_1$ by Eqs.~\eqref{am1}--\eqref{am4}. The cost function $f_2$ can be expressed entirely in terms of the independent variables $w_{0x}^\ast$, $w_{0y}^\ast$, $w_{0z}^\ast$, as follows:
\[
  f_2= (w_{0x}-w_{0x}^\ast)^2+(w_{0y}-w_{0y}^\ast)^2+(w_{0z}-w_{0z}^\ast)^2+
  (w_{1x}^\ast-w_{0x}^\ast)^2+(w_{1y}^\ast-w_{0y}^\ast)^2+(w_{1z}^\ast-w_{0z}^\ast)^2.
\]
The critical points can be found by setting the partial derivatives of $f_2$ with respect
to $w_{0x}^\ast$, $w_{0y}^\ast$, $w_{0z}^\ast$ to zero and solving the resulting system of
equations. Doing so, only one solution appears:
\[
  w_{0x}^\ast = w_{1x} = \frac{w_{0x}+w_{1x}^\ast}{2}, \;
  w_{0y}^\ast = w_{1y} = \frac{w_{0y}+w_{1y}^\ast}{2}, \;
  w_{0z}^\ast = w_{1z} = \frac{w_{0z}+w_{1z}^\ast}{2}.
\]
In the other case, i.e. $\hat{\vec{r}}_0=-\hat{\vec{r}}_1=(1,0,0)$, the values of $k_0$ and $k_1$ are
not necessarily equal, hence $\vec{r}_0=(k_0^{-1},0,0)$ and $\vec{r}_1=(-k_1^{-1},0,0)$, but $w_1$ can
be expressed in terms of $w_0^\ast$.
Indeed, the conservation laws for the angular momentum $\vec{h}$ and eccentricity vector $\vec{e}$ in the intermediate orbit imply that $\vec{r}_0\times\vec{w}_0^\ast=\vec{r}_1\times\vec{w}_1$ and $\vec{w}_0^\ast\times(\vec{r}_0\times\vec{w}_0^\ast)-\hat{\vec{r}}_0=
 \vec{w}_1\times(\vec{r}_1\times\vec{w}_1)-\hat{\vec{r}}_1$, respectively, from which it follows that:
\[
  w_{1x} = w_{0x}^\ast, \quad
  w_{1y} = -\frac{k_1}{k_0}w_{0y}^\ast, \quad
  w_{1z} = -\frac{k_1}{k_0}w_{0z}^\ast.
\]
The cost function $f_2$ can be written in terms of the independent variables $w_{0x}^\ast$, $w_{0y}^\ast$, $w_{0z}^\ast$, as follows:
\[
  f_2= (w_{0x}-w_{0x}^\ast)^2+(w_{0y}-w_{0y}^\ast)^2+(w_{0z}-w_{0z}^\ast)^2+
  (w_{1x}^\ast-w_{0x}^\ast)^2+\left(w_{1y}^\ast+\frac{k_1}{k_0}w_{0y}^\ast \right)^2+\left(w_{1z}^\ast+\frac{k_1}{k_0}w_{0z}^\ast \right)^2.
\]
Taking partial derivatives and solving the resulting system of equations, we get the unique solution:
\[
  w_{0x}^\ast = \frac{w_{0x}+w_{1x}^\ast}{2}, \;
  w_{0y}^\ast = k_0\frac{k_0w_{0y}-k_1w_{1y}^\ast}{k_0^2+k_1^2}, \;
  w_{0z}^\ast = k_0\frac{k_0w_{0z}-k_1w_{1z}^\ast}{k_0^2+k_1^2}.
\]

\section{Optimality of the Hohmann transfer}\label{sec-hohmann}

In this problem, we want to find the optimal 2-impulse transfer between concentric
and coplanar circular orbits. Assuming that the plane that contains both initial
and final orbits is orthogonal to $(0,0,1)$ and that the initial point is on the
$x$-axis, we can reduce to the following situation:
\[
    \vec{l}_0 = (0,0,l_{0z}), \;
    \vec{l}_2 = (0,0,l_{2z}), \;
    \vec{s}_0 = \vec{s}_2 = (0,0,0), \;
    \hat{\vec{r}}_0 = (1,0,0),
\]
where $l_{0z}$ and $l_{2z}$ are not zero.

The nine unknowns are the components of the vectors $\vec{l}_1=(l_{1x},l_{1y},l_{1z})$, $\vec{s}_1=(s_{1x},s_{1y},s_{1z})$ and $\hat{\vec{r}}_1=(x_1,y_1,z_1)$.
The seven equations relating them are:
\[
\left\{
  \begin{aligned}
    & l_{1x}s_{1x}+l_{1y}s_{1y}+l_{1z}s_{1z} = 0 \\
    & l_{1x} = 0 \\
    & l_{1x}x_1+l_{1y}y_1+l_{1z}z_1 = 0 \\
    & l_{2z}z_1 = 0 \\
    & x_1^2+y_1^2+z_1^2 = 1 \\
    & l_{0z}^2 = l_{1x}^2+l_{1y}^2+l_{1z}^2+s_{1y}l_{1z}-s_{1z}l_{1y} \\
    & l_{2z}^2 = l_{1x}^2+l_{1y}^2+l_{1z}^2+x_1(s_{1y}l_{1z}-s_{1z}l_{1y})+ y_1(s_{1z}l_{1x}-s_{1x}l_{1z})+z_1(s_{1x}l_{1y}-s_{1y}l_{1x})
  \end{aligned}
  \right.
\]
Since $l_{2z}\neq 0$, we have that $z_1=0$. Substituting $l_{1x}=0$ and $z_1=0$ in the third
equation, we get $l_{1y}y_1=0$. We will discuss two cases: $l_{1y}=0$ and $l_{1y}\neq0$ (which means that $y_1=0$).

\bigskip

Case $l_{1y}=0$. Here we can reduce to $\vec{l}_1=(0,0,l_{1z})$, $\vec{s}_1=(s_{1x},s_{1y},0)$ and $\hat{\vec{r}}_1=(x_1,y_1,0)$, subject to the following conditions:
\[
\left\{
  \begin{aligned}
    & x_1^2 + y_1^2 = 1 \\
    & l_{0z}^2 = l_{1z}^2 + s_{1y}l_{1z} \\
    & l_{2z}^2 = l_{1z}^2 + x_1s_{1y}l_{1z}-y_1s_{1x}l_{1z}
  \end{aligned}
  \right.
\]
Since we want to minimize the function $f_1$, we need to introduce the two extra variables $\Delta_0$
and $\Delta_1$, which represent the magnitude of each impulse and the following two restrictions:
\[
\left\{
  \begin{aligned}
    \Delta_0^2 &= s_{1x}^2 + (s_{1y}+l_{1z}-l_{0z})^2 \\
    \Delta_1^2 &= (s_{1x}-l_{1z}y_1+l_{2z}y_1)^2 + (s_{1y}+l_{1z}x_1-l_{2z}x_1)^2
  \end{aligned}
  \right.
\]
Using Lagrange multipliers as in Eq.\eqref{eq-alfa}, we computed the reduced Gr\"obner basis in the polynomial ring $\mathbb{Q}(l_{0z},l_{2z})[\lambda_1,\ldots,\lambda_5,\Delta_0,\Delta_1,s_{1x},
s_{1y},l_{1z},x_1,y_1]$ with respect to the lexicographic monomial order $\lambda_1>\cdots>\lambda_5>\Delta_1>\Delta_0>s_{1x}>
s_{1y}>l_{1z}>x_1>y_1$, obtaining the following two solutions:
\[
    x_1 = -1, \;
    y_1 = 0, \;
    l_{1z} = \pm\sqrt\frac{l_{0z}^2+l_{2z}^2}{2}, \;
    s_{1x} = 0, \;
    s_{1y} = \frac{l_{0z}^2-l_{2z}^2}{l_{0z}^2+l_{2z}^2}l_{1z} \, .
\]
The cost function $f_1=\Delta_0+\Delta_1$ at each of those solutions becomes:
\begin{equation}\label{eq-fuel-hohmann}
  f_1 = \left|\frac{\pm\sqrt{2}l_{0z}^2}{\sqrt{l_{0z}^2+l_{2z}^2}}-l_{0z} \right| +
   \left|\frac{\pm\sqrt{2}l_{2z}^2}{\sqrt{l_{0z}^2+l_{2z}^2}}-l_{2z} \right|.
\end{equation}
A simple computation shows that the sign of the optimum $l_{1z}$ (and also the sign of the numerators in the previous expression) coincides with the sign of $l_{0z}+l_{2z}$. This case corresponds to the classical Hohmann solution.

\bigskip

Case $l_{1y}\neq 0$. Since $l_{1y}y_1=0$ and $x_1^2+y_1^2=1$, this case is only possible when
$y_1=0$ and $x_1=\pm 1$.

When $x_1=1$, we have $l_{0z}^2=l_{2z}^2$, so the the orbits are of the same radius.
If $l_{0z}=l_{2z}$, the initial and final orbits are exactly the same and no maneuver is needed.
On the other hand, if $l_{0z}=-l_{2z}$, the satellite must change the direction of rotation in
two impulses, both at the same point, so $\hat{\vec{r}}_0=\hat{\vec{r}}_1$ and
$\vec{w}_0^*=\vec{w}_1$. There are clearly infinitely many optimal solutions with $f_1=\Delta_1+\Delta_2=2|\vec{w}_0|$.

It only remains the case $x_1=-1$. Here the unknowns are $\vec{l}_1=(0,l_{1y},l_{1z})$ and
$\vec{s}_1=(s_{1x},s_{1y},s_{1z})$, subject to the equations
\[
\left\{
  \begin{aligned}
    & l_{1y}s_{1y}+l_{1z}s_{1z} = 0 \\
    & l_{0z}^2 = l_{1y}^2+l_{1z}^2+s_{1y}l_{1z}-s_{1z}l_{1y} \\
    & l_{2z}^2 = l_{1y}^2+l_{1z}^2-s_{1y}l_{1z}+s_{1z}l_{1y}.
  \end{aligned}
  \right.
\]
The impulses are
\[
\left\{
  \begin{aligned}
    \Delta_0^2 &= s_{1x}^2 + (s_{1y}+l_{1z}-l_{0z})^2 + (s_{1z}-l_{1y})^2 \\
    \Delta_1^2 &= s_{1x}^2 + (s_{1y}-l_{1z}+l_{2z})^2 + (s_{1z}+l_{1y})^2 \\
  \end{aligned}
  \right.
\]
and the cost function is $f_1=\Delta_0+\Delta_1$. We computed the Gr\"obner basis of Eq.\eqref{eq-alfa} in the ring $\mathbb{Q}(l_{0z},l_{2z})[\lambda_1,\ldots,\lambda_5,\Delta_0,\Delta_1,s_{1x},s_{1y},s_{1z},l_{1y},l_{1z}]$ with respect to the lexicographic monomial order $\lambda_1>\cdots >\lambda_5 >\Delta_1 >\Delta_1 > s_{1x} > s_{1y} >s_{1z} >l_{1y} >l_{1z}$, obtaining the following two optimal solutions:
\begin{align*}
l_{1z}&=\frac{l_{0z}^5+l_{0z}^4 l_{2z}+4l_{0z}^3l_{2z}^2+4l_{0z}^2 l_{2z}^3 + l_{0z} l_{2z}^4+l_{2z}^5}{4l_{0z}l_{2z} (l_{0z}^2+l_{0z}l_{2z}+l_{2z}^2)}\\
l_{1y}&=\pm \sqrt{\frac{l_{0z}^2+l_{2z}^2}{2}-l_{1z}^2}\\
s_{1z}&=-\frac{l_{0z}^2-l_{2z}^2}{l_{0z}^2+l_{2z}^2} \, l_{1y}\\
s_{1y}&=\frac{l_{0z}^2-l_{2z}^2}{l_{0z}^2+l_{2z}^2} \, l_{1z}\\
s_{1x}&=0
\end{align*}

These solutions are defined only when $a_1 <\frac{l_{2z}}{l_{0z}}<a_2$, where $a_1$ and $a_2$ are the real roots of $a^4+2a^3+2a+1=0$. In particular, the sign of $\frac{l_{2z}}{l_{0z}}$ must be negative. Substituting these solutions in the cost function $f_1$ and comparing with  Eq.\eqref{eq-fuel-hohmann}, it can be checked that the solution of the previous case is always better.

\section{Two rotated ellipses}\label{sec-two-rot}

In this orbit-to-orbit transfer problem, we will assume that the initial and final orbits are two
identical ellipses rotated an angle $\alpha\in(0,\pi]$ lying on the same plane. We will restrict
our optimization to intermediate orbits that also lie within the same plane, i.e. to a two-dimensional orbit
transfer problem. Without loss of generality, we can assume that $\vec{l}_0=(0,0,l_{0z})$, $\vec{l}_2=(0,0,l_{2z})$, $\vec{s}_0=(s_{0x},s_{0y},0)$, $\vec{s}_2=(s_{2x},s_{2y},0)$ are given, and that we have to find $\hat{\vec{r}}_0=(x_0,y_0,0)$, $\hat{\vec{r}}_1=(x_1,y_1,0)$, $\vec{l}_1=(0,0,l_{1z})$ and $\vec{s}_1=(s_{1x},s_{1y},0)$. In order to guarantee that the initial and final orbits have the same eccentricity and semi-major axis, and to maximize the symmetry of the equations, we impose $l_{2z}=l_{0z}=1$ (since the problem does not depend on the semi-major axis), $s_{2x}=-s_{0x}$ and $s_{2y}=s_{0y}$. This ensures that the orbits are identical, but rotated an angle $\alpha=2\arctan(s_{0x}/s_{0y})$. Both orbits are also symmetric with respect to the $x$-axis. Finally, we need two extra variables $\Delta_0$ and $\Delta_1$ to represent the two impulses.

For general ellipses, with arbitrary semi-latus rectum $p$, all the values $\vec{l}_1$, $\vec{s}_1$, $\vec{w}_0^*$, $\vec{w}_1$, $\Delta_0$ and $\Delta_1$ calculated in this section have to be divided by $\sqrt{p}$.

The discussion above reduces the problem to two parameters $s_{0x}$, $s_{0y}$, nine unknowns $x_0$, $y_0$, $x_1$, $y_1$,
$s_{1x}$, $s_{1y}$, $l_{1z}$, $\Delta_0$, $\Delta_1$, six equations
\[
\left\{
\begin{aligned}
 eq_1:=&x_0^2+y_0^2=1\\
 eq_2:=&x_1^2+y_1^2=1 \\
 eq_3:=&l_{1z}^2+l_{1z}(x_0s_{1y}-y_0s_{1x})-1-x_0s_{0y}+y_0s_{0x} =0\\
 eq_4:=&l_{1z}^2+l_{1z}(x_1s_{1y}-y_1s_{1x})-1-x_1s_{0y}-y_1s_{0x} =0\\
 eq_5:=&\Delta_0^2 = (s_{0x}-s_{1x})^2+(s_{0y}-s_{1y})^2+(1-l_{1z})^2+2(1-l_{1z})(x_0(s_{0y}-s_{1y})-y_0(s_{0x}-s_{1x})) \\
 eq_6:=&\Delta_1^2 = (s_{0x}+s_{1x})^2+(s_{0y}-s_{1y})^2+(1-l_{1z})^2+2(1-l_{1z})(x_1(s_{0y}-s_{1y})+y_1(s_{0x}+s_{1x}))
\end{aligned}
\right.
\]
and a cost function $f_1=\Delta_0+\Delta_1$.

After introducing the Lagrange multipliers, the algebraic problem has $15$ equations and $15$ unknowns.
Although such a system is expected to have a finite number of solutions, this is not true in our problem,
so some special treatment is needed. We will divide the problem in several cases, which will be discussed below.

\bigskip

Case 1: We impose the extra condition $y_0+y_1\neq 0$, which is done algebraically by introducing an additional variable $k$ and
adding the equation $1-k(y_0+y_1)$ to the system. Geometrically, this new system looks for orbit transfers that are not symmetric with respect to the $x$-axis. We have no proof that the system has always a finite number of solutions, but we have collected extensive numerical evidence that this is indeed true. The best orbit transfer never
happened to come from this case, as shown in Subsection~\ref{sec-num}.

\bigskip

Case 2: Now we consider the remaining case, i.e. $y_0+y_1=0$. It follows from $eq_1$ and $eq_2$ that
$x_1=\pm x_0$, so we split the analysis again: $x_0=x_1$ (case 2a) and $x_0=-x_1$ (case 2b). The former
represents transfers whose initial and final points are symmetric with respect the $x$-axis, and the latter
is a degenerate case when the initial point, the final point and the origin are collinear. Both cases have a finite number of solutions, which we will compute explicitly below.

\bigskip

Case 2a: We assume here that $x_1=x_0$ and $y_1=-y_0$. Subtracting $eq_4$ from $eq_3$, we obtain that
$y_0s_{1x}=0$. When $y_0=0$, we have $x_0=x_1=\pm 1$ and the cost function can be written as $f_1=\sqrt{(s_{1x}-s_{0x})^2+A}+\sqrt{(s_{1x}+s_{0x})^2+A}$,
where $A$ is an expression that does not involve $s_{1x}$. Since $s_{1x}$ vanishes from all the equations
when $y_0=0$, we can consider it as a free variable. Setting the derivative of $f_1$ with respect to $s_{1x}$ to zero and solving the equation gives $s_{1x}=0$ after some algebraic manipulation. Now substituting $x_0=x_1=\pm 1$, $y_0=y_1=0$, $s_{1x}=0$ in the equations, leaves us with only one restriction $l_{1z}^2\pm l_{1z} s_{1y} \mp s_{0y}=1$. Solving for $s_{1y}$ and substituting everything in the cost function $f_1$, we get a minimization problem with only $l_{1z}$ as a free variable, which gives the following two solutions:
\begin{equation}\label{sol-caso2a-ast}
  x_0=x_1=\pm 1,\;
  y_0=y_1=0,\;
  s_{1x}=0,\;
  s_{1y}=s_{0y},\;
  l_{1z}=1,\;
  f_1=2|s_{0x}|.
\end{equation}
It only remains to see what happens when $s_{1x}=0$ and $y_0\neq 0$. Comparing $eq_5$ and $eq_6$ shows
that $\Delta_0=\Delta_1$, so we can replace the cost function $f_1$ by \[
\frac{1}{2}f_2=\Delta_0^2=s_{0x}^2+(s_{0y}-s_{1y})^2+(1-l_{1z})^2+2(1-l_{1z})(x_0(s_{0y}-s_{1y})-y_0s_{0x}).
\]
This reduces the problem to four
unknowns $x_0$, $y_0$, $s_{1y}$, $l_{1z}$, subject to two equations $eq_1$ and $eq_3$. The case $x_0=0$ leads easily
to the following two solutions:
\begin{equation}\label{sol-caso2a-t2}
\begin{aligned}
  &x_0 = x_1 = 0,\;
  y_0 = \pm 1,\;
  y_1 = \mp 1,\;
  s_{1x} = 0,\;
  s_{1y} = s_{0y},\;
  l_{1z} = \sqrt{1\mp s_{0x}},\\
  &f_1 = 2| 1\mp s_{0x} \mp \sqrt{1\mp s_{0x}}|.
\end{aligned}
\end{equation}
A straightforward verification shows that $|s_{0x}|>\ 2| 1\mp s_{0x} \mp \sqrt{1\mp s_{0x}}|$
for all $s_{0x}\in(-1,1)$, which means that the solution~\eqref{sol-caso2a-t2} is always better than~\eqref{sol-caso2a-ast}.

From now on, we assume that $x_0\neq 0$. This allows us to express $s_{1y}$ in terms of $x_0$, $y_0$, $l_{1z}$ using $eq_3$, as follows:
\[
  s_{1y}=\frac{1+x_0s_{0y}-y_0s_{0x}-l_{1z}^2}{l_{1z}x_0}
\]
Substituting the expression for $s_{1y}$ in the cost function $\frac12 f_2$, we obtain a rational function $c(x_0,y_0,l_{1z})$. Therefore, we have to minimize $c$ subject to $x_0^2+y_0^2-1=0$, which is
equivalent to solving the equations
\[
  \left\{
  \begin{aligned}
  	\frac{\partial c}{\partial l_{1z}} =0\\
  	 x_0\frac{\partial c}{\partial y_0}
  	 	-y_0\frac{\partial c}{\partial x_0} =0\\
  	 x_0^2+y_0^2-1 = 0
  \end{aligned}
  \right.
\]
We can clear denominators, without losing any information, by multiplying the first equation by $l_{1z}^3x_0^2$ and the second one by $l_{1z}^2x_0^3$, since $l_{1z}$ and $x_0$ are both non-zero. We define the equations
\[
\begin{aligned}
  eq_7 := &l_{1z}^3x_0^2\frac{\partial c}{\partial l_{1z}}=0\\
  eq_8 := &l_{1z}^2x_0^3\left(x_0\frac{\partial c}{\partial y_0}-y_0\frac{\partial c}{\partial x_0}\right)=0
\end{aligned}
\]
so our system is equivalent to solving $eq_1$, $eq_7$ and $eq_8$. To solve these algebraic equations, we use
resultants. Taking advantage of the fact that $eq_1$ does not contain any term involving $l_{1z}$, we define
$p_{7,8}:={\rm Res}_{l_{1z}}(eq_7,eq_8)\in\mathbb{Z}(s_{0x},s_{0y})[x_0,y_0]$ and $p_{1,7,8}:={\rm Res}_{x_0}(eq_1, p_{7,8})\in\mathbb{Z}(s_{0x},s_{0y})[y_0]$. Any solution of our system satisfies both $p_{7,8}$ and $p_{1,7,8}$. Conversely,
a solution of $p_{1,7,8}=0$ can be extended to a solution
of the original system $\{eq_1,eq_7,eq_8\}$ using the following procedure:
\begin{enumerate}
 \item Find a solution $y_0\in\mathbb{R}$ of $p_{1,7,8}(y_0)=0$.
 \item Write $p_{7,8}(x_0,y_0)=\sum_{i\geq 0}a_i(y_0)x_0^i\in\mathbb{Z}(s_{0x},s_{0y})[y_0][x_0]$ and substitute every
  even power $x_0^{2j}$ by $(1-y_0^2)^j$ and every odd power $x_0^{2j+1}$ by $x_0(1-y_0^2)^j$. This way we
  obtain a polynomial of degree one in $x_0$ with coefficients in $\mathbb{Z}(s_{0x},s_{0y})[y_0]$. This substitution is
  correct since $x_0$ satisfies $eq_1=x_0^2+y_0^2-1=0$.
 \item Assume that the polynomial obtained in the previous step is $q_1(y_0)x_0+q_0(y_0)$. Then, calculate
 $x_0=-q_0(y_0)/q_1(y_0)\in\mathbb{R}$.
 \item Apply the Euclidean algorithm to the polynomials of $eq_7$ and $eq_8$ until a polynomial of degree one in $l_{1z}$
  appears.
 \item Assume that the polynomial obtained in the previous step is $r_1(x_0,y_0)l_{1z}+r_0(x_0,y_0)$. Then,
  calculate $l_{1z}=-r_0(x_0,y_0)/r_1(x_0,y_0)$.
\end{enumerate}

The theory of resultants (see for instance Section~3 of \cite{CLO05}) guarantees that the polynomials  $q_1(y_0)$ and $r_1(x_0,y_0)$ are different from zero. Since $p_{1,7,8}$ is a polynomial
of degree $48$, there are potentially $48$ different solutions to our system of equations. A closer
inspection of $p_{1,7,8}$ shows that it factorizes as
\[
  y_0^8(y_0-1)^6(y_0+1)^6(y_0^2s_{0x}^2-2y_0s_{0x}+y_0^2s_{0y}^2+1-s_{0y}^2)^4pol_{20}(y_0),
\]
where $pol_{20}(y_0)\in\mathbb{Z}(s_{0x},s_{0y})[y_0]$ is an irreducible polynomial of degree $20$. Therefore, the roots of $p_{1,7,8}$ are $0$ with multiplicity $8$, $1$ and $-1$ with multiplicity $6$, the complex numbers
\[
  \frac{s_{0x}\pm s_{0y}\sqrt{s_{0x}^2+s_{0y}^2-1}}{s_{0x}^2+s_{0y}^2}
\]
with multiplicity $4$, and the $20$ different roots of $pol_{20}(y_0)=0$. The roots $0$, $1$, $-1$ are
discarded since we have already excluded these subcases. The complex roots with multiplicity $4$ are
not real since $s_{0x}^2+s_{0y}^2$ is the square of the eccentricity, which is always $<1$, by assumption.
This leaves only the $20$ roots of $pol_{20}(y_0)$ to be considered.

Note also that the polynomials $pol_{20}$,
$q_0$, $q_1$, $r_0$ and $r_1$ can be precomputed symbolically (as explained above), so the solution is:
\begin{equation}\label{sol-caso2a-t1}
\begin{aligned}
  &y_0=-y_1=\text{a root of}\;pol_{20},\;
  x_0=x_1=-\frac{q_1(y_0)}{q_0(y_0)},\;
  s_{1x}=0,\\
  &s_{1y}=\frac{1+x_0s_{0y}-y_0s_{0x}-l_{1z}^2}{l_{1z}x_0},\;
  l_{1z}=-\frac{r_1(x_0,y_0)}{r_0(x_0,y_0)}.
\end{aligned}
\end{equation}
All together, we have $22$ different solutions of case 2a: $2$ from Eq.~\eqref{sol-caso2a-t2} and $20$ from Eq.~\eqref{sol-caso2a-t1}.
Extensive numerical evidence shows
that the best transfer always comes from one of these solutions, as discussed in Subsection~\ref{sec-num}.

\bigskip

Case 2b: We assume here that $y_1=-y_0$ and $x_1=-x_0$. In this case, we have $\frac{1}{2}(eq_3+eq_4)=l_{1z}^2-1+y_0s_{0x}=0$, which implies that $y_0=\frac{1-l_{1z}^2}{s_{0x}}$. In the particular
case when $l_{1z}=1$, the following two solutions are found directly from the equations:
\begin{equation}\label{sol-caso2b-ast}
  x_0=\pm 1,\; x_1=\mp 1,\;
  y_0=y_1=0,\;
  l_{1z}=1,\;
  s_{1y}=s_{0y},\;
  s_{1x}\in[-|s_{0x}|,|s_{0x}|],\;
  f_1=2|s_{0x}|.
\end{equation}
When $l_{1z}=-1$, then only one solution is possible:
\begin{equation}\label{sol-caso2b-cuad}
\begin{aligned}
  &x_0=1,\; x_1=-1,\;
  y_0=y_1=0,\;
  s_{1y}=-s_{0y},\;
  s_{1x}=-s_{0x}s_{0y},\\
  &l_{1z}=-1,\;
  f_1=\left(|1+s_{0y}|+|1-s_{0y}|\right)\sqrt{4+s_{0x}^2} \, .
\end{aligned}
\end{equation}
The solution~\eqref{sol-caso2b-cuad} has $f_1\geq 2\sqrt{4+s_{0x}^2}>2|s_{0x}|$, so it is always
worse that~\eqref{sol-caso2b-ast} and can be discarded. It can also be shown that solution~\eqref{sol-caso2b-ast} is always worse than~\eqref{sol-caso2a-t2}, so it can be safely ignored as well.

\bigskip

The only remaining case is $|l_{1z}|\neq 1$. Using $eq_3-eq_4$, we can write $s_{1x}$ in terms of
the other unknowns:
\[
  s_{1x} = \frac{x_0(l_{1z}s_{1y}-s_{0y})s_{0x}}{l_{1z}(1-l_{1z}^2)}.
\]
The problem has now only three variables $x_0$, $s_{1y}$ and $l_{1z}$ and a single constraint $eq_1$, which after substituting $y_0=\frac{1-l_{1z}^2}{s_{0x}}$ becomes:
\[
  eq_9:=s_{0x}^2(x_0^2-1)+(1-l_{1z}^2)^2=0 \, .
\]
The cost function $f_1=\sqrt{\Delta_0^2}+\sqrt{\Delta_1^2}$ is the sum of the square roots of two
rational expressions in $x_0$, $s_{1y}$ and $l_{1z}$. In this case, we will not introduce the extra
variables $\Delta_0$ and $\Delta_1$, since the square roots can be removed with an algebraic trick.
First, according to the theory of Lagrange multipliers, we should have $\nabla f_1=\lambda \nabla eq_9$
at each local extrema of $f_1$. This produces the following equations:
\begin{equation} \label{eq44}
\left\{
\begin{aligned}
  \frac{1}{2\sqrt{\Delta_0^2}}\frac{\partial\Delta_0^2}{\partial s_{1y}}+
  \frac{1}{2\sqrt{\Delta_1^2}}\frac{\partial\Delta_1^2}{\partial s_{1y}}&=0  \\
  \frac{1}{2\sqrt{\Delta_0^2}}\frac{\partial\Delta_0^2}{\partial x_0}+
  \frac{1}{2\sqrt{\Delta_1^2}}\frac{\partial\Delta_1^2}{\partial x_0}&=2\lambda s_{0x}^2x_0 \\
  \frac{1}{2\sqrt{\Delta_0^2}}\frac{\partial\Delta_0^2}{\partial l_{1z}}+
  \frac{1}{2\sqrt{\Delta_1^2}}\frac{\partial\Delta_1^2}{\partial l_{1z}}&=-4\lambda l_{1z}(1-l_{1z}^2)
\end{aligned}
\right.
\end{equation}

We can remove $\lambda$ by multiplying the last two equations of \eqref{eq44} by $2l_{1z}(1-l_{1z}^2)$ and $s_{0x}^2x_0$, respectively, and adding them.
\begin{equation}
  \frac{1}{2\sqrt{\Delta_0^2}}\left(
    2l_{1z}(1-l_{1z}^2)\frac{\partial\Delta_0^2}{\partial x_0}+s_{0x}^2x_0\frac{\partial\Delta_0^2}{\partial l_{1z}}\right)+
  \frac{1}{2\sqrt{\Delta_1^2}}\left(
    2l_{1z}(1-l_{1z}^2)\frac{\partial\Delta_1^2}{\partial x_0}+s_{0x}^2x_0\frac{\partial\Delta_1^2}{\partial l_{1z}}\right)=0
    \label{eq47}
\end{equation}
Finally, we remove the square roots in the first equation of Eq.~\eqref{eq44} and also in \eqref{eq47} by moving one of the terms to the right, and then squaring both sides. After some algebraic manipulation and the introduction of the non-zero factors $l_{1z}^3(l_{1z}+1)^2(l_{1z}^2-1)^2$ and $l_{1z}^6(l_{1z}+1)(l_{1z}^2-1)^5$ to clear the denominators, we get the following two polynomials:
\[
\begin{aligned}
  eq_{10}&:=l_{1z}^3(l_{1z}+1)^2(l_{1z}^2-1)^2\left[\left(\frac{\partial\Delta_0^2}{\partial s_{1y}}\right)^2\Delta_1^2-\left(\frac{\partial\Delta_1^2}{\partial s_{1y}}\right)^2\Delta_0^2\right]=0\\
  eq_{11}&:=l_{1z}^6(l_{1z}+1)(l_{1z}^2-1)^5\left[\left(
    2l_{1z}(1-l_{1z}^2)\frac{\partial\Delta_0^2}{\partial x_0}+s_{0x}^2x_0\frac{\partial\Delta_0^2}{\partial l_{1z}}\right)^2\Delta_1^2 \right.\\
    &\left. \qquad \qquad \qquad \qquad \qquad \qquad -\left(
    2l_{1z}(1-l_{1z}^2)\frac{\partial\Delta_1^2}{\partial x_0}+s_{0x}^2x_0\frac{\partial\Delta_1^2}{\partial l_{1z}}\right)^2\Delta_0^2 \right]=0
\end{aligned}
\]
At this point we have reduced the whole case 2b to three polynomial equations $\{eq_9,eq_{10},eq_{11}\}$ in three unknowns $x_0$, $s_{1y}$ and $l_{1z}$. To solve the system, we exploit the
fact that $eq_9$ does not contain the variable $s_{1y}$. Define $p_{10,11}:={\rm Res}_{s_{1y}}(eq_{10},eq_{11})\in\mathbb{Z}(s_{0x},s_{0y})[x_0,l_{1z}]$ and $p_{9,10,11}:={\rm Res}_{x_0}(eq_9,p_{10,11})\in\mathbb{Z}(s_{0x},s_{0y})[l_{1z}]$.

A similar procedure to the one used in case 2a allows us to obtain a full solution of the equations $\{eq_9,eq_{10},eq_{11}\}$ from a zero of $p_{9,10,11}$. The procedure is described below:
\begin{enumerate}
  \item Calculate a root $l_{1z}\in\mathbb{R}$ of $p_{9,10,11}=0$.
  Although $p_{9,10,11}$ is a polynomial of degree $166$ in $l_{1z}$, it can be factored as a product of polynomials that are either not zero by assumption or that have degree lower than $6$.
  \item Use $eq_{9}$ to obtain the two possible values of $x_0$:
   \[
      x_0=\pm\sqrt{1-\left(\frac{1-l_{1z}^2}{s_{0x}}\right)^2} \, .
   \]
      If these values are not real numbers, then the following
      steps can be skipped.
  \item Apply the Euclidean algorithm to the polynomials of $eq_{10}$ and $eq_{11}$ to compute their greatest common divisor. The algorithm stops when a polynomial of degree $1$ in $s_{1y}$ appears.
  \item Assume that the polynomial obtained in the previous step is $u_1(x_0,l_{1z})s_{1y}+u_0(x_0,l_{1z})$. Then calculate $s_{1y}=-\frac{u_0(x_0,l_{1z})}{u_1(x_0,l_{1z})}$.
  \item The value of the remaining variables are:
  \[
     x_1=-x_0,\; y_0=-y_1=\frac{1-l_{1z}^2}{s_{0x}},\;
     s_{1x}=\frac{x_0(l_{1z}s_{1y}-s_{0y})s_{0x}}{l_{1z}(1-l_{1z}^2)} \, .
  \]
\end{enumerate}
The polynomials $p_{9,10,11}$, $u_0$ and $u_1$ can be precomputed symbolically, following the same procedure as above. These polynomials can therefore be reused to calculate the solutions of case 2b for any given $s_{0x}$ and $s_{0y}$.
\begin{equation}\label{sol_caso2b}
\begin{aligned}
  &l_{1z} = \text{a root of}\;p_{9,10,11},\;
  x_0 = -x_1 = \pm\sqrt{1-\left(\frac{1-l_{1z}^2}{s_{0x}}\right)^2},\;
  y_0 = -y_1 = \frac{1-l_{1z}^2}{s_{0x}},\\
  &s_{1y} = -\frac{u_0(x_0,l_{1z})}{u_1(x_0,l_{1z})},\;
  s_{1x} = \frac{x_0(l_{1z}s_{1y}-s_{0y})s_{0x}}{l_{1z}(1-l_{1z}^2)}.
\end{aligned}
\end{equation}
We have collected extensive numerical evidence showing that these solutions are always worse than those of case 2a. Anyways, since there are only a finite number of solutions in this case, which can be computed by the explicit formula Eq.~\eqref{sol_caso2b}, we recommend that these solutions are included when looking for the best orbit transfer.

\subsection{Numerical tests} \label{sec-num}

In our numerical computations, we explored a wide range
of values of $s_{0x}$ and $s_{0y}$, in such a way that all possible
eccentricities and angles between the ellipses were considered.

When the ellipses are rotated 180 degrees, the optimal solution is always provided by Eq.~\eqref{sol-caso2a-t2} of case2a.

Indeed, case1 does not have a real solution if the eccentricity is $0.1, 0.2, \ldots, 0.9$. Solutions of case2b given by Eq.~\eqref{sol_caso2b} and case2a given by Eq.~\eqref{sol-caso2a-t1} exist but are worse.

In the rest of the cases,  the solution of case2a given by Eq.~\eqref{sol-caso2a-t1} is always the best. To check this efficiently, we used only rational values for $s_{0x}$ and $s_{0y}$. The trick to achieve this is to set
\[
  s_{0x} = e\frac{a^2-b^2}{a^2+b^ 2},\; s_{0y}=e\frac{2ab}{a^2+b^2}
\]
where $e$ is the desired eccentricity and $a$ and $b$ are integers
chosen in such a way that the desired angle $\alpha$ is approximately $2\arctan\left(\frac{a^2-b^2}{2ab}\right)$.

In the following tests, we used $e=0.1,0.2,\ldots,0.9$ and the
pairs $(a,b)$ were selected to approximate the angles $\alpha=5,10,\ldots,175$ degrees. The case $\alpha=180$ was discussed above.

For each value of $a$, $b$ and $e$, we computed the best solution
of case 1 (solving the system numerically), case2a using both Eq.~\eqref{sol-caso2a-t1} and Eq.~\eqref{sol-caso2a-t2}, and case2b using Eq.~\eqref{sol_caso2b}. In total, we have explored more than a thousand test cases. We extracted several conclusions from the data we computed.

\begin{figure}[H]
\centering
\begin{minipage}[c]{0.45\textwidth}
 \centering
\includegraphics[width=\textwidth]{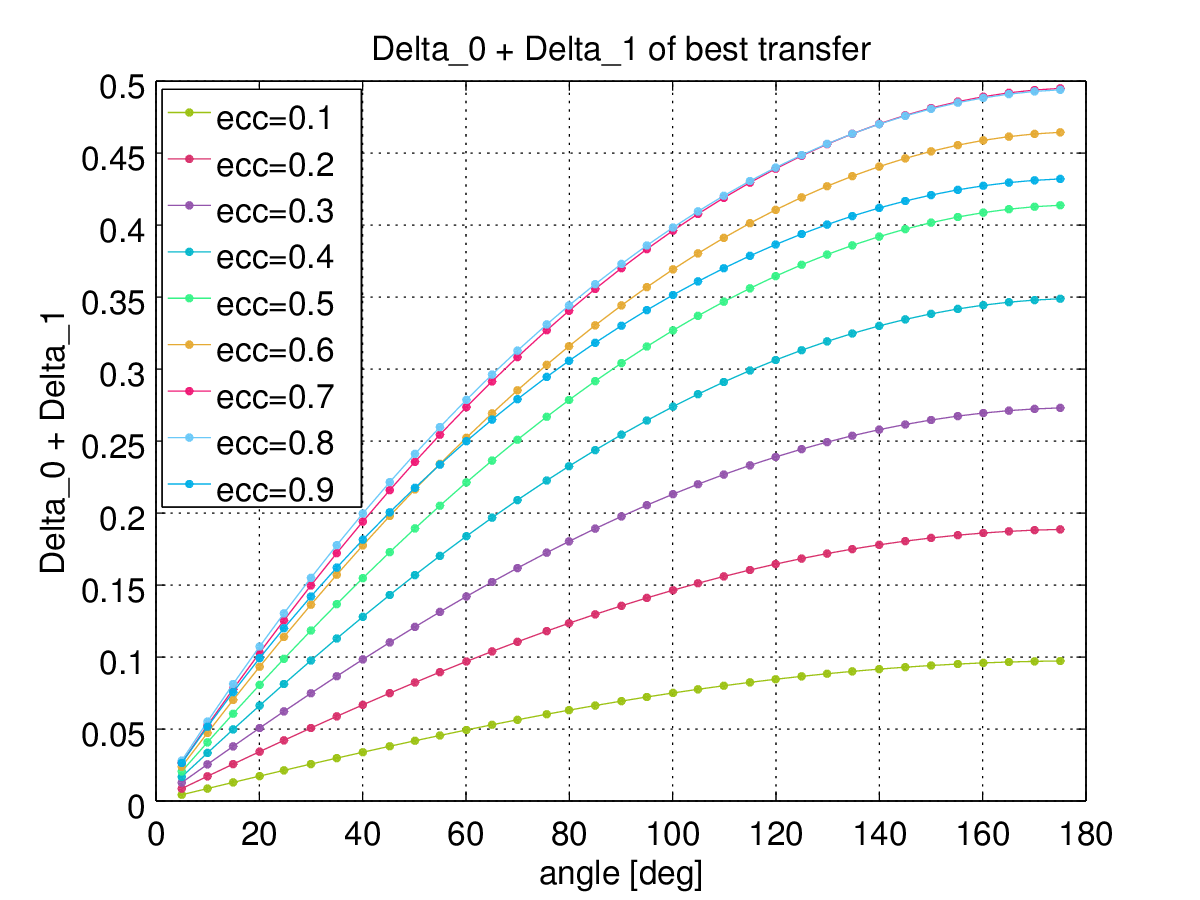}
\end{minipage}
\begin{minipage}[c]{0.45\textwidth}
 \centering
\includegraphics[width=\textwidth]{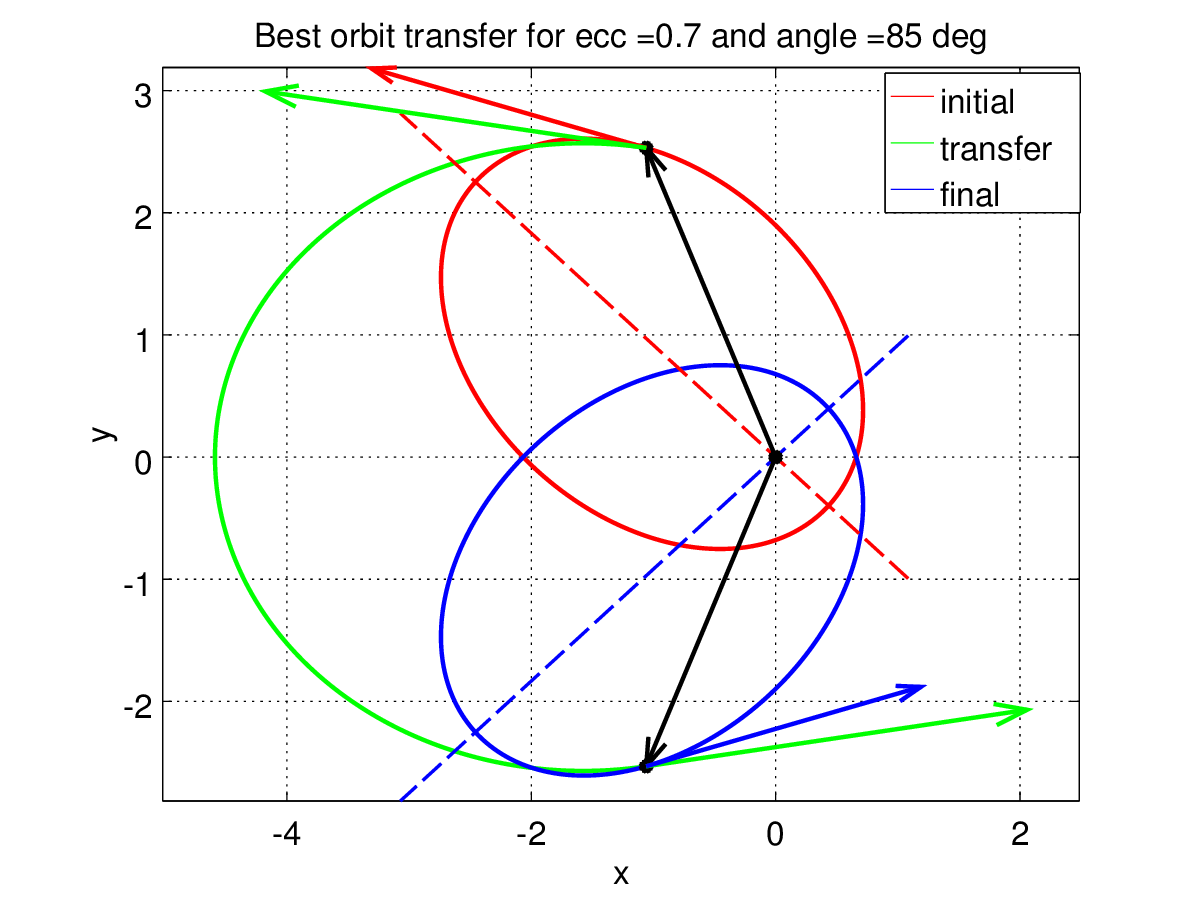}
\end{minipage}
\caption{$\Delta_0+\Delta_1$ of the best transfer and an example of one of these transfers.}
\label{fig-caso2at1}
\end{figure}

First of all, the solution of case2a is indeed the best one, as we mentioned before. In Figure~\ref{fig-caso2at1} we show the fuel of this transfer (or rather, the cost function $f_1=\Delta_0+\Delta_1$, which is proportional to it) obtained as a function of $e$ and $\alpha$, and what this transfer would look like when $e=0.7$ and the angle between the ellipses is $85$ degrees.

\begin{figure}[H]
\centering
 \begin{minipage}[c]{0.45\textwidth}
 \centering
\includegraphics[width=\textwidth]{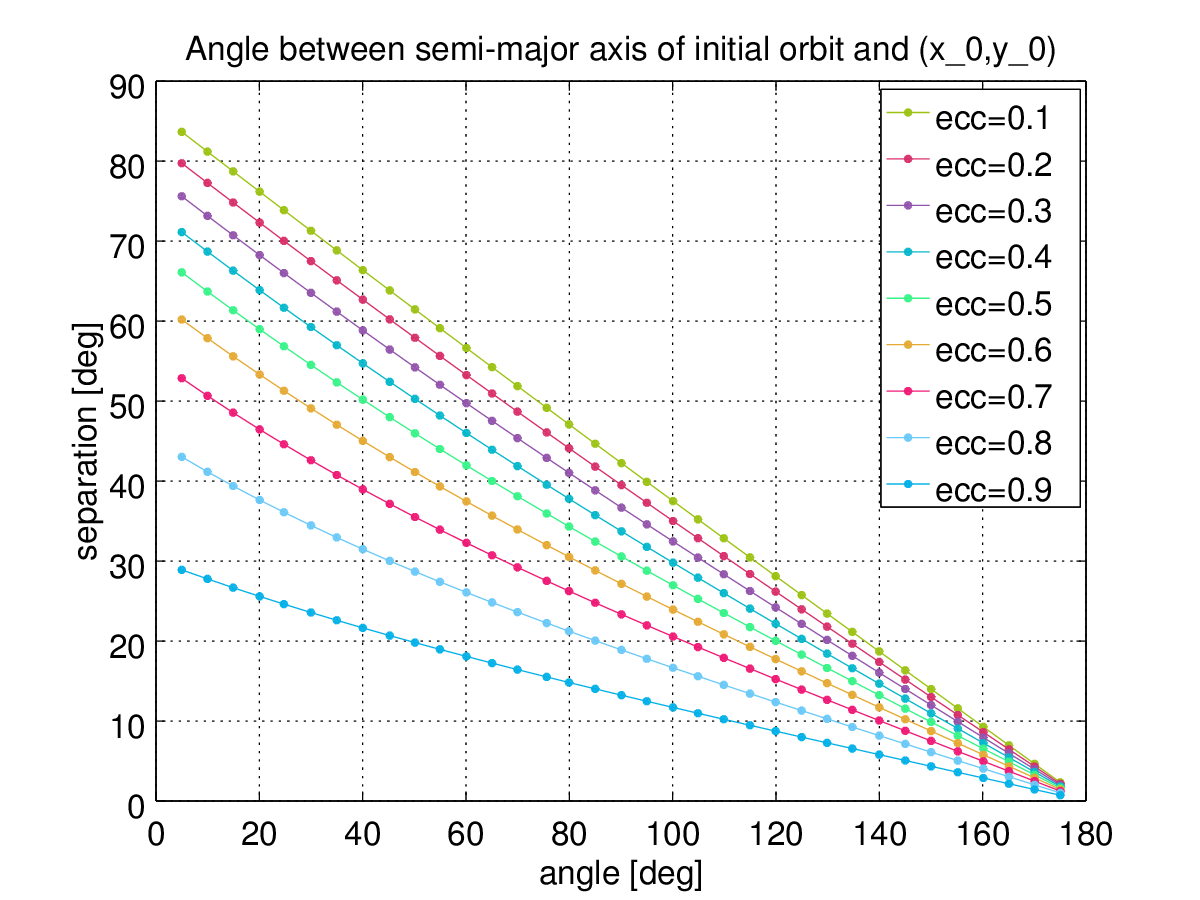}

(a)
\end{minipage}
\begin{minipage}[c]{0.45\textwidth}
\centering
\includegraphics[width=\textwidth]{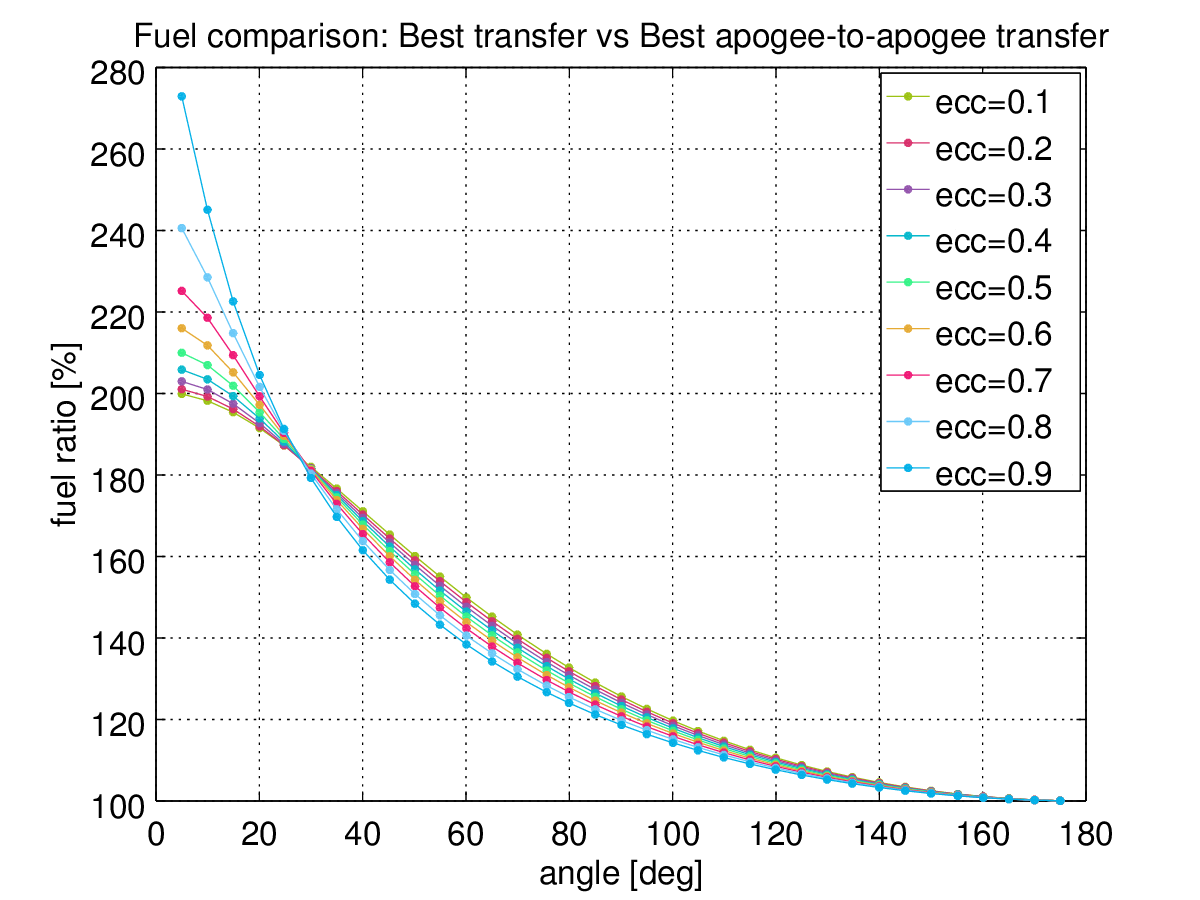}

(b)
\end{minipage}
\caption{(a) Angle of separation between the semi-major axis of the initial orbit and the point where the first impulse is applied. (b) Fuel comparison between the best transfer and the best one from apogee to apogee.}
\label{fig-separacion}
\end{figure}

We can also compare the angle between the semi-major axis of the initial orbit and the direction given by $\vec{r}_0$. We observe in Figure~\ref{fig-separacion}(a) that when $\alpha$ is small (less than 40 degrees), this separation is higher than 50 degrees for eccentricities up to 0.5. When $\alpha$ is near 180 degrees, this separation becomes smaller (and is zero in the case $\alpha=180$).

The separation shown above led us to study how much fuel can be saved by using our optimal transfer instead of the one from apogee to apogee. Figure~\ref{fig-separacion}(b) shows the ratio (in percentage) between the fuel consumption of both transfers.

On the other hand, case1 does not always produce a valid real solution. Even in those situations where case1 provides a solution, it is always very poor compared to the ones of case2a, up to one or two orders of magnitude worse depending on the eccentricity. Case2b always produces valid solutions, but they are as bad as those of case1.

\begin{figure}[H]
\centering
 \begin{minipage}[c]{0.6\textwidth}
 \centering
\includegraphics[width=\textwidth]{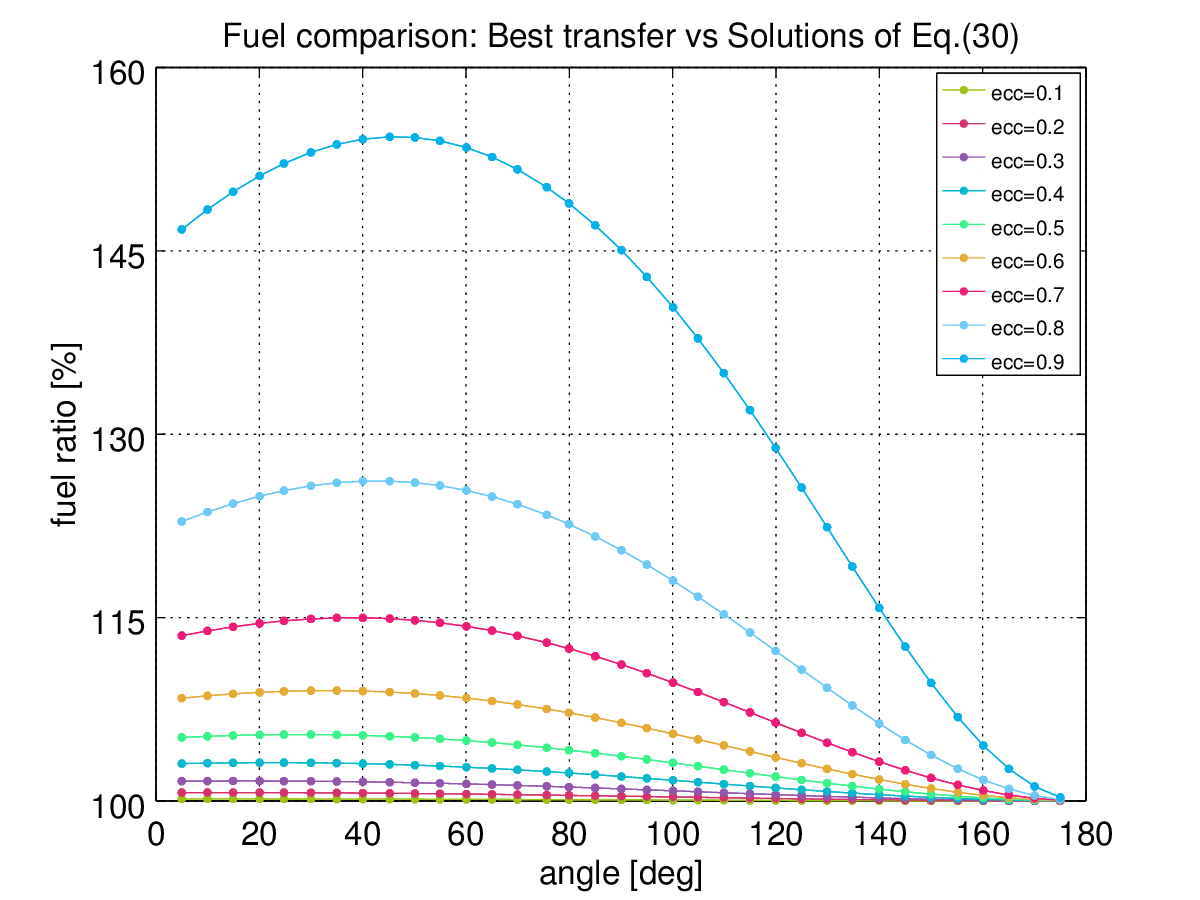}
\end{minipage}
\caption{Fuel comparison between the best transfer and the solutions provided by Eq.\eqref{sol-caso2a-t2}}
\label{fig-caso2at1vscaso2at2}
\end{figure}

Finally, the solutions of case2a given by Eq.~\eqref{sol-caso2a-t2} are worse than the optimal one, but they are no more than $10\%$ worse for eccentricities below $0.6$ and up to a $55\%$ worse for higher eccentricities, as shown in Figure~\ref{fig-caso2at1vscaso2at2}.

\section{Conclusions}

Firstly, in Section~\ref{sec-modelo} we have presented an innovative approach to
the study of the multi-impulse orbit transfer problem with minimum fuel,
that only requires to solve a system of polynomial equations. No trigonometrical functions are needed. This allows one to use all the algebraic machinery that is currently available: Gr\"obner bases, Resultants, Elimination theory, Root Isolation, etc., which was not
possible with the previous formulations of the problem.

In Section~\ref{sec-am} we provided an alternative method for solving the
point-to-point two-dimensional orbit transfer problem studied in~\cite{AM10} by Avenda\~no and Mortari. The significant advantage
of our new approach is the efficiency of the computation, since only
arithmetic operations are used.

Moreover, in Section~\ref{sec-hohmann} we used the well-known Hohmann transfer
problem between two circular orbits, to show the power of our technique.
We worked under very general assumptions, i.e. letting the transfer orbit
to be out of plane, but we showed that the best transfer is indeed
coplanar. The novelty of our analysis is that we allowed the initial and final orbits to have angular momentum pointing in opposite directions. Even for such an extreme case, the classical solution is proven optimal.

Furthermore, in Section~\ref{sec-two-rot} we analyzed in depth the problem of changing
between two identical elliptical orbits of eccentricity $e$ which are coplanar and rotated a certain angle $\alpha$. Here we restricted our search to transfer orbits that
are coplanar with the other two. The first surprising result that we got is that the optimal transfer does not go from apogee to apogee (except when $\alpha=180\,{\rm deg}$), but it is separated from the apogee a certain variable angle that depends on~$e$ and~$\alpha$. This angle is higher than $50$ degrees when $e\leq 0.5$ and $\alpha\leq40$ deg. For lower eccentricities and small values of $\alpha$, the separation angle can be as large as $80$ degrees.

The large difference between the best transfer orbit and the best one from apogee to apogee means also a significant difference in the
fuel consumption of both maneuvers. For angles $\alpha$ up to $80$ degrees, the savings obtained by using our transfer orbit are always higher than $25\%$, and for angles $\alpha\leq 10\,\text{deg}$, our transfer consumes less than half of the fuel needed to go from apogee to apogee.

Finally, solving the problem has been reduced to the study of several subproblems, each of which consists of a set of polynomial equations.
All but one of the subproblems have been solved symbolically, which means that an explicit solution is available given
any initial and final orbits. The remaining subproblem can be solved numerically for any given data, but the optimal solution does not seem to come from this particular subcase. Indeed, numerical evidence suggests that the optimal solution always comes from the same subcase, for which we have a symbolic solution.

\section*{Acknowledgements}

The first author is partially supported by the MINECO grant ESP2013-44217-R, the second author by the MINECO grant MTM2011-22621 and the FQM-327 group (Junta de Andaluc\'{i}a, Spain) and the third author by the MINECO grant MTM2013-45710-C2-1-P and the groups E15 Geometr\'{i}a (DGA, Spain) and FQM-333 (Junta de Andaluc\'{i}a, Spain). The last three authors are also partially supported by the ``Centro Universitario de la Defensa de Zaragoza'' grant ID2013-15.

\end{document}